\documentstyle{camera}

\begin{document}

\title{PHENOMENOLOGICAL QUANTUM GRAVITY: THE BIRTH OF A NEW FRONTIER?}

\author{R. Aloisio$^a$, P. Blasi$^b$, A. Galante$^c$, P.L. Ghia$^d$,
  A.F. GRILLO$^a$
and F. Mendez$^a$}
\organization{$^a$INFN - Laboratori Nazionali del 
Gran Sasso, SS. 17bis,
Assergi (L'Aquila) - Italy\\
$^b$INAF - Osservatorio Astrofisico di Arcetri, 
Largo E. Fermi 5
50125 Firenze - Italy\\
$^c$Dipartimento di Fisica, 
Universit\`a di L'Aquila, Via Vetoio
67100 Coppito (L'Aquila) - Italy\\
$^d$CNR - IFSI, Sezione di Torino, Corso Fiume 4, 
10133 Torino - Italy, and INFN - Sezione di Torino, 
Via P. Giuria 1, 10125 Torino - Italy}

\maketitle

\begin{abstract}
In the last years a general consensus has emerged that, contrary to
intuition, quantum-gravity effects may have relevant consequences for the 
propagation and interaction of high energy particles. This has given
birth to the field of ``Phenomenological Quantum Gravity'' We review
some of the aspects of this new, very exciting frontier of Physics.
\end{abstract}

\section{Introduction}
In the late 50's John Weeler \cite{wheel} made clear that when gravity 
(described by general relativity) is coupled to quantum mechanics, the 
concept of space-time itself changes: in fact space-time becomes dynamical 
and when examined at very  small distances, near 
$l_P=\sqrt(hc/G_N^3)\approx 10^{-33}~cm$ ($G_N$ being the Newton constant 
and $l_P$ is called the ``Planck distance'') has to show violent fluctuations 
making for instance impossible to define a distance. The emergence of these 
phenomena has been generically named {\it space-time foam}, but their effects 
were confortably thought to be visible only in processes not testable in 
laboratory physics.

However the Universe has more surprises than we might expect. 

One of the consequences of general relativity (and in fact historically
preceeding it) is that in flat space-time, as is approximately ours at
least for distances much smaller than the scale of the Universe,
Lorentz Invariance should hold. This has been tested with high
precision. It is however important to remind that relativistic
invariance stems from  experimental facts and as such has to be put
under scrutiny. It is in fact expected that in {\it space-time foam} 
regime violations of relativistic invariance might appear.
So (minuscule) departures from relativistic invariance will in general 
signal the onset of the QG regime.

And in fact Nature provides us with  very
sensitive tools to test Lorentz Invariance: the Cosmic Microwave
Background Radiation (CMBR) and other universal radiations. On
this radiation UHE cosmic ray protons can interact and loose energy
quite efficiently and it is expected that their spectrum bends at the
highest energies. It is the famous Greisen, Zatsepin and Kuzmin (GZK)
cut-off \cite{GZK}, with a threshold at $E \approx 5~10^{19}~eV$. The
energy lost by protons in the CMBR is such that they can only travel
for about $100~Mp$, before being brought below the threshold. So the
spectrum should show a sharp decrease above the threshold, the amount
being related to the distribution and evolution of possible sources.

Notwistanding the extreme energy of the threshold, the process involved 
is a low energy one, and in fact one of the best known experimentally: 
the pion photoproduction $\gamma p \to N \pi$  whose cross-section is 
extremely well known, in the frame in which the target proton is at rest. 
This absorption appears at extreme energies just because it is the process 
above, but examined in a reference frame where the photon has an extremely
 low energy ($\approx 10^{-3}~eV$). It is therefore clear that a
verification of the presence of the GZK cut-off would imply a
verification of Lorentz boosts between frames with a Lorentz factor
$\gamma_L \approx 5\cdot 10^{10}$.

This fact was discovered already few years
after the theoretical determination of the GZK cut-off, in 1971  in a paper by 
Kirshinitz and Chechin \cite{K&C}, which went largely unnoticed. At
the time there were already CR experiments (Haverah Park, Yakutsk) with
the possibility of detecting CR primaries at these energies, but the
situation was not clear. In their pioneering work  they wrote 
{\it ``Primary protons with energy above $5\cdot 10^{19} eV$ are expected to
be strongly slowed down by the interaction with the background
thermal radiation. However, no break is observed in the CR
spectrum in this region. It is of course premature in this
circumstances....''} and the key observation was
{\it ``The point is that the primary
protons have a uniquely large Lorentz factor $\gamma > 5 10^{10}$
larger by many order of magnitudes than in any other experiment..''}.

With these premises they proposed a modified relativity theory to introduce 
small violations in the dispersion relation of particles at sufficiently high 
energies in such a way to account for the absence of the so-called GZK feature 
in the spectrum. This may be taken as the official date of birth of 
``phenomenological quantum gravity'' although it might not have been named 
in such way those days. 

The point, as we will see in a moment, is that there are processes in
which possible tiny violations of Lorentz invariance, normally
absolutely negligible, can be in some way amplified by peculiar
physical situations (here,  the extreme Lorentz factor between the
frames in which the GZK effect is expected and the laboratory in which
cross sections are measured), generally typical of astrophysical
contexts, in such a way to induce in principle measurable
effects. Theories in which relativistic invariance is modified are then
in general falsifiable, and in fact many classes of them have already been.

More than 30 years after the experimental situation concerning UHECRs is still 
unclear: previous largest experiments, AGASA \cite{AGASA}
and HiRes \cite{Hires} do not provide strong evidence either 
in favor or against the detection of the GZK feature \cite{demarco}. 
A substantial increase in the statistics of events, as expected 
with the Auger project \cite{Auger}  will
clarify the scenario in one or two more years. 

\section{Relativistic invariance modifications from Quantum Gravity}

As anticipated above, there is {\it a priori} no guarantee that, when
space-time becomes dynamical, relativistic invariance is preserved
down to the smallest distances (or highest energies).

Several attempts to construct a model for QG
have been done. They basically share a new interpretation for
space-time: it is no more a given background for physical objects
These attempts include Loop QG, some string-based model and the 
space-time foam approach.

However the status of QG theories, although much more evolved than
even a few years ago, does not still allow to describe a realistic 
low (compared to the Planck) energy limit where the effects of
modification of relativistic invariance should be experimentally 
verifiable/falsifiable.

Therefore several  models have
been proposed as effective theories that should try to catch some
of the possible new QG physics at large but still sub-Planckian
energy scales. 

All these approaches predict some
modification of basic physical principles. The following is
a non-exhaustive cumulative list of the different possibilities.
The first is the possibility of modification of Poincar\'e and
Lorentz symmetries. Depending on the specific model they can 
still be exactly realized as well as 
explicitly broken (introducing a preferred reference frame)
or kept but in a deformed way. The energy-momentum (dispersion)
relation is generally  modified including extra terms that can be of
fixed or stochastic nature.  
Generally a new invariant physical scale ($l_p$ or the Planck
energy $E_p$) is introduced and this scale can (possibly)
coexist with the standard invariant: 
the (low energy limit of) light speed $c$, which may in fact acquire
an energy dependence.
Other possible effects are indetermination in position and/or
momentum measurements due to the fluctuating nature of the 
space-time structure and the appearance of new non-linear 
composition laws for energy and momentum of multiparticle 
states $i.e.$ $P_{tot}\neq \sum_i P_i$.

Many of these possibilities have been investigated
trying to find possible experimental signatures for new
physics even at energy scales much smaller the $10^{28}$ eV
that correspond to the Plank energy.

Astroparticle physics is a privileged arena for such studies 
both for the availability of very energetic particles
and for the possibility to consider their motion along large 
(cosmological) distances.
Among the others the large distance propagation of photons
with energy dependent velocity \cite{am97,biller99,steck01} 
and modifications induced in 
the standard synchrotron radiation emission process have been
considered to put limits on possible Lorentz Invariance (LI) 
breaking \cite{jac03,ellis03,castorina}.

Another interesting possibility to test such models is to
consider  physical processes with a kinematic energy threshold, 
which is in turn very sensitive to the smallest violations of LI.
This is the case for UHECRs and TeV gamma rays.
UHECRs are expected to suffer severe 
energy losses due to photopion production off the photons of the cosmic 
microwave background (CMB), and this should suppress the flux of particles 
at the Earth at energies above $\sim 10^{20}$ eV, the so called GZK feature. 
Super-TeV energy photons from sources at cosmological distances
are expected to undergo electron-positron production in interactions
with low energy photons of the far infra red background (FIRB) and 
CMB.

In both cases a very large $\gamma$ factor is involved in moving
from the laboratory to the center of mass reference frame.
The sharply defined thresholds can be substantially shifted 
(or even disappear) if a small LI breaking term is introduced
giving potential for investigation in this field.
The new phenomena, if present,  
should show up in modification of expected UHECRs
spectrum.

Some authors \cite{cam,colgla,berto,steck04}
have invoked possible violations of LI as a plausible explanation to
some puzzling observations related to the detection of ultra high energy
cosmic rays (UHECRs) with energy above the GZK feature, 
and to the unexpected shape of the spectrum of photons with super-TeV 
energy from sources at cosmological distances.
 
Both types of observations have in fact 
many uncertainties, either coming from limited statistics of very rare events,
or from accuracy issues in the energy determination of the detected 
particles, and most likely the solution to the alleged puzzles will come from 
more accurate observations rather than by a violation of fundamental 
symmetries.
 
For this reason, from the very beginning we proposed \cite{noi1} that 
cosmic ray observations should be used as an ideal tool to constrain 
the minuscule violations of LI, rather than as evidence for the need 
to violate LI.

We adopt  some reasonable choice to parametrize the LI violations
predicted by QG models, consider the theoretical consequences 
and compare with experimental data.
If the features in the spectrum related to the processes thresholds 
are indeed found this will provide limits on LI violation scale.
If such features are absent this will allow us to reject some
models but, for the moment, not to prove the existence of 
LI breaking new phenomena.

\section{More on Lorentz invariance modifications}

The recipes for the modifications of LI generally consist of requiring 
modification (either explicit or stochastic) of the dispersion
relation  of high energy  particles.
This modification is an effective way to describe 
their propagation in the ``vacuum'', now affected by 
quantum gravity (QG) phenomena. 
This effect is generally parametrized by introducing 
a mass scale $M$, expected to be of the order of the Planck mass,
that sets the scale for QG to become effective. 

Without referring to any specific model, we write a modified 
dispersion relation obeying the following postulates:

1) 
modifications are universal, $i.e.$ do not depend on
particle type;

2) 
modifications preserve rotational invariance;

3) 
modifications are an high energy phenomenon, vanishing at  low 
momenta.
\\
\noindent
With these conditions we write the following expression:

\begin{equation}
E^2-p^2=\mu(m,p/M)=m^2+p^2f(p/M)
\label{disp_rel}
\end{equation}
This deformed dispersion relation has been proposed by several
authors \cite{luk98,ellis99,DSR1,DSR2} and is the most popular in the
literature.

Just for completeness we
note that another possibility compatible with the dimensional
analysis exists: it refers to the so called conformal models of
LI modifications and was considered by Kirzhnits and Chechin in their
paper. It accounts to introduce the extra (respect to the standard
case) term proportional to the particle mass squared instead that
to $p^2$. When considering thresholds modifications this last 
possibility gives no detectable effects for UHECRs propagation
if $M$ is the Planck mass.

Already at this stage we can intuitively understand why modification
of the dispersion relations can sensitively affect the threshold
values: the right-hand side of the modified dispersion relation can be
thought as a (momentum dependent) effective mass, and the thresholds
do depend explicitely on rest masses: we therefore expect strong
modifications.

The standard way to proceed is to expand the last term in rhs
of (\ref{disp_rel}) and this, at lowest order, gives a term of the
form 
\begin{equation}
E^2-p^2=m^2+\eta (p/M)^\alpha
\label{extra}
\end{equation}
where $\alpha$ is model dependent and $\eta$ a real parameter of order
one.
To get a quick result and some physical insight
we can argue that, for massive particles, the above extra term in
dispersion relation becomes relevant for the kinematics of
particle interactions
when its modulus is comparable with the particle squared mass.
For the protons ($i.e.$ for the GZK case) 
we get immediately the following numbers for the critical momentum
$p_c$ where we may expect changes (in the following formula we fix
$M$ to the Planck mass value):

\begin{eqnarray}
\alpha=1 &\to & p_c=(m_p^2 M^2)^\frac{1}{3}\simeq 10^{15} eV << M \nonumber\\
\alpha=2 &\to & p_c=(m_p^2 M^2)^\frac{1}{4}\simeq 10^{18} eV << M \nonumber
\end{eqnarray}

In both case we see that the value of $p_c$ is much smaller than the
Planck mass scale, justifying $a~posteriori$ the Taylor expansion. 
This gives another indication that if we modify
the dispersion relation with terms related to some scale (the Planck
mass in our case),
the resulting particle kinematics can indeed be sensitive to such 
changes already at much lower energy scales. In other words  
 we do not need Planck scale experiments to detect effects
related to new physics at Planck scale.

A detailed calculation of photopion and $e^+e^-$ threshold production 
for high energy protons and photons interacting with low energy
background photons has been carried out \cite{noi1}. In this calculation
the conservation of total energy and momentum of incoming and outcoming
particles is assumed. 

If the total energy and momentum of multiparticle states are calculated
as usual (just the sum of the contribution of each particle) and we assume
that the scale parameter $M$ is the Planck mass we find that the 
GZK feature could be absent (the threshold goes to infinity) when
we consider $\eta $ negative, or, for positive
sign, shifted downward by five ($\alpha=1$) or one ($\alpha=2$) order 
of magnitude respect to the standard case.

Notice that (as an aside) the same equations that do describe the
modifications of the thresholds also imply that for positive sign some
decay processes like $\gamma \to e^+ e^-$ can happen at high energies
and this severely restricts the range of allowed modification
parameters. From a theoretical point of view notice this means that
physics might be different in different reference frames, as expected
when Lorentz symmetry is violated.

The same calculations can be done in the framework of Doubly Special 
Relativity. In this case the theory is constructed in such a way 
that the relativity principle is still valid: no privileged 
reference system exists.
The (non linear) deformed boost in
momentum space require
a change in the dispersion relation as the one previously considered
but also a different definition of total energy and momentum in multiparticle
states. For the DSR1 \cite{DSR1} and DSR2 \cite{DSR2} models we have
(at lowest order in $p/M$) 
\cite{judes04}:
\begin{eqnarray}
E_{tot} = E_1 + E_2 - \frac{1}{2}\frac{1}{M} (p_1 p_2 + p_2 p_1) + 
O(\frac{1}{M^2}) \nonumber\\
E_{tot} = E_1 + E_2 + \frac{1}{M} (E_1 E_2 + E_2 E_1) + 
O(\frac{1}{M^2}) \nonumber
\end{eqnarray}
In this case basically no new particle processes (like photon decay)
are kinematically allowed and, for the GZK case,
the momentum threshold is basically the same as in standard case
\cite{major04}.

In drawing conclusions from this kind of studies we have to keep in
mind that there are two main problems. The first is related to the up
to now relatively poor and conflicting experimental data on UHECR
spectrum. This will be discussed in more detail in the next
section. The second is related to the limitation of approaches based uniquely
on kinematic analysis: the present impossibility to include 
the dynamical effects of the full theory makes quantitative 
conclusions questionable (even if it seems reasonable to 
expect modifications to dynamics to be proportional to the energy scale
divided by $M$ and hence highly suppressed for physics below
GZK scale).

We conclude this section by remarking that it is also possible to
assume that modifications are of stochastic nature, i.e. $\eta$
in Eq. 2 becomes a random variable. In fact we can think that this is
more natural since we do expect that the space-time itself shows
fluctations near the Planck scale, and that modifications as the ones
discussed above emerge as the result of some averaging over quantum
fluctations. Many aspects of this approach are discussed in the
contribution of J.Y. Ng to this workshop \cite{vulcng}. Here we want
only remark that, when propagation becomes stochastic, in general both
signs of $\eta $ are possible, and in fact unavoidable. This has many
striking consequences, as discussed in detail in \cite{noi3}, the most
unexpected being that $all$ charged particles do emit photons in the
vacuum at all energies, in principle loosing catastrophically energy. 
 
In this case the threshold can be written as
\begin{equation}
p_{th} \simeq \left(\frac{m^2 M \omega}{\delta}\right)^{\frac{1}{4}}
\end{equation}
where $m$ is the particle mass, 
$\omega$ is the photon energy and $\delta$ is some combination
of  fluctuating coefficients.
Clearly $p_{th} \to 0$ if $\omega \to 0$ and this will eventually
result in a stability crisis for all charged particles \cite{noi3,gallou04}.
It is not clear at present how to avoid this problem, but certainly
this will set conditions on allowable theories.

\section{The Ultra High Energy Cosmic Ray spectrum: present state and
  perspectives for Phenomenological  Quantum Gravity}

The Cosmic Ray spectrum at and above the GZK cut-off is presently
known with large statistical and systematic errors; moreover the two
largest experiments till now have presented conflicting evidence
(although at less than $3~\sigma$ level): AGASA does not find
evidence of the cut-off while the spectrum presented by HiRes shows
the expected decrease.

A new generation of experiments is developping and already the Pierre
Auger Observatory (P.A.O, see \cite{tiina} in these proceedings),
still in its building phase, is the largest experiment in the
world. When the flux will be delivered at the $2005$ summer conferences, a
statistic analogous to the total AGASA statistic will be employed,
although systematic errors are likely to be still relatively large.

Then in a year or two the question of the existence of a sharp decrese
of the spectrum will be settled and we will be left with only two
possibilities: 
\begin{itemize}
\item The bend in the spectrum will be found at the expected position,
  as in the HiRes data: then some kind of relativistic invariance must
  hold at least up to $10^{20}~eV$, like in normal Lorentz Invariant
  theory, as well as in most DSR approaches\footnote{At least one
  example exists of a $ad~hoc$ DSR theory that produces large shift of
  thresholds, without violating frame independence
  \cite{amegrande}.}. It is in general difficult to distinguish
  between the two approaches. In some DSR flavours, however, real photons can
  acquire a energy-dependent speed, and this can be tested in future
  satellite experiments like GLAST. Finally, since we do not know the
  sources of Cosmic Rays at these energies, but we know that
  acceleration becomes less and less efficient at high energies, we
  cannot exclude that the budget of UHECRs sources is simply
  vanishing. In principle this possibility can be tested if the
  spectrum at energies sub-GZK is known with large precision, a
  measurement that  P.A.O. can accomplish. 
\item No bending will be found. It is tempting to conclude that this
  will signal the onset of new physics connected to quantum gravity
  and violations of relativistic invariance. However this is at
  present unjustified, the reason still being that we do not know the
  sources at these energies: several alternative possibilities
  exist. However the sources should be relatively nearby
  $(D<100~Mp)$. If (with the statistic allowed by P.A.O.) a
  statistically significant correlation of super-GZK events with very
  distant sources will be found, than propagation in the Universe (and
  therefore relativistic invariance) will be at a question.
\end{itemize}

As a final speculation, we remark that if it were possible to detect
the spectrum of a single source, an exponential decreas (rather then a
bend) is expected. This can be done with photons. Some controversial claim has
already been reported at tens of TeV energies, where absorption is on
the FIRB, whose spectrum is  however  affected by systematics. The
same measurement on the CMBR would imply detection of PeV photons,
with the $bonus$ that the interaction length becomes smaller than the
galactic radius. So the detection of even a single extragalactic (say,
in the Magellanic Clouds) PeV source, although terribly demanding from an
experimental point of view, would unambiguously signal departure from
relativistic invariant propagation \cite{berber}.


\begin{thebibliography}{9}
\bibitem{berber}This point was raised to us some years ago by V. Berezinsky.
\end{thebibliography}
\end{document}